\documentclass[aps,pra,twocolumn,groupedaddress,showpacs]{revtex4}
\usepackage{graphicx}
\usepackage{amsmath}

\begin{document}

\title{Fidelity of Optically Controlled Single- and Two-Qubit Operations on Coulomb-Coupled Quantum Dots}



\author{Juliane Danckwerts}\email{juliane@physik.tu-berlin.de}
\author{Andreas Knorr}
\author{Carsten Weber}
\affiliation{Institut f\"ur Theoretische Physik, Nichtlineare Optik und Quantenelektronik, Technische Universit\"at Berlin, Hardenbergstr. 36, 10623 Berlin}

\date{September 7, 2010, published as Phys. Status Solidi B 247, 2147 (2010)}





\pacs{73.21.La, 03.67.Lx, 03.67.Bg} 

%
%
%

\begin{abstract}
We investigate the effect of the Coulomb interaction on the applicability of quantum gates on a system of two Coulomb-coupled quantum dots. We calculate the fidelity for a single- and a two-qubit gate and the creation of Bell states in the system. The influence of radiative damping is also studied. We find that the application of quantum gates based on the Coulomb interaction leads to significant input state-dependent errors which strongly depend on the Coulomb coupling strength. By optimizing the Coulomb matrix elements via the material and the external field parameters, error rates in the range of $10^{-3}$ can be reached. Radiative dephasing is a more serious problem and typically leads to larger errors on the order of $10^{-2}$ for the considered gates. In the specific case of the generation of a maximally entangled Bell state, error rates in the range of $10^{-3}$ can be achieved even in the presence of radiative dephasing.
\end{abstract}

%
%

\maketitle   

\section{Introduction}

Quantum dots (QDs) have long been discussed as a source for semiconductor-based quantum information processing \cite{Loss:PhysRevA:98}. For a physical realization of a quantum computer, a quantum mechanical two-level system that can serve as a qubit is required. This system must allow for preparation, manipulation, and readout of its quantum state \cite{Nielsen::00}. Furthermore, a coupling mechanism between the qubits is needed to allow for conditional operations on multiple qubits. Other interactions, especially damping mechanisms that affect the two-level system, have to be minimized.

As representations of qubits, excitonic states in single or double QDs as well as spin states have been proposed \cite{Lovett:PhysRevB:03,Lovett:PhysRevB:05,Biolatti:PhysRevB:02,Chen:PhysRevB:04,Reiter:PhysRevLett:09}. Quantum information schemes in QD systems are often 
based on the coherent optical control of the electronic states, taking advantage of the femtosecond time scales common in current ultrafast laser optics \cite{Chen:PhysRevB:04,Reiter:PhysRevLett:09}. The Coulomb interaction is one of the coupling mechanisms that has been proven to allow for conditional operations on QD qubits  \cite{Biolatti:PhysRevLett:00}. However, it also influences the optical manipulation of the QD states, especially for small interdot distances where the Coulomb interaction between neighbouring QDs is an important coupling mechanism \cite{Lovett:PhysRevB:03,Unold:PhysRevLett:05,Richter:PhysStatusSolidiB:06}.
Therefore, it affects the applicability of any quantum information scheme based on QDs, even if the scheme does not utilize the Coulomb interaction for conditional operations. Here, we want to quantify this influence and discuss the implications for QD based quantum information processing.

The key quantity to describe successful quantum information processing in a physical system is the fidelity with which gates can be implemented on the considered set of qubits \cite{Nielsen::00}. For a basic evaluation of a given proposal for quantum information processing, the fidelity of a universal set of gates, consisting, e.g., of single-qubit gates and the CNOT gate, has to be investigated. Fidelity calculations have been presented for single-qubit gates \cite{Biolatti:PhysRevB:02,Caillet:EurPhysJD:07,Grodecka:PhysRevB:07} and two-qubit gates based on spin states in single \cite{Emary:PhysRevB:07} and coupled quantum dots \cite{Xu:PhysRevLett:09} as well as excitonic states in single QDs \cite{Piermarocchi:PhysRevB:02}. The fidelity of two-qubit gates in Coulomb-coupled QDs, especially  with entangled output states,  has not been discussed in the literature.
 
In this article, we present calculations of the fidelity of a single- and a two-qubit gate implemented on Coulomb-coupled QDs. We focus on a scheme utilizing the Coulomb interaction for the implementation of a controlled two-qubit gate, namely the CNOT gate \cite{Biolatti:PhysRevB:02,Lovett:PhysRevB:03}. The quantum operations are implemented using coherent light pulses resonant to the excitonic transitions of the system. By applying a series of coherent laser pulses with fixed areas, special quantum information gates  can be realized. In particular, we consider the generation of Bell states in this system.

\section{Hamiltonian}

We consider a system consisting of two QDs, interacting with a coherent light field and with each other via the Coulomb interaction. It is assumed that the wave functions of the electrons in different dots do not overlap, i.e., we neglect interdot tunneling. 
Even though we focus on the low temperature regime (a few Kelvin), dephasing effects due to the interaction with phonons can be of importance, depending 
strongly on the pulse length and detuning of the dot levels \cite{Grodecka::05,Grodecka:PhysRevB:07}. Since the resulting errors are typically an order of magnitude smaller than the one arising from radiative dephasing, which is studied in this work, we do not considered electron-phonon coupling in the following.

Following the proposition in Ref. \cite{Biolatti:PhysRevLett:00}, the qubits are represented by the single-excitonic states in each QD, one state being the QD ground state, the other being the lowest exciton 
state (exciton ground state). 
The reason is as follows: Due to the exchange interaction, the two heavy hole and two electronic states closest to the Fermi edge typically show fine structure splitting into two bright and dark excitonic states each, which are energetically split depending on the symmetries of the QDs \cite{Bayer:PhysRevB:02,Seguin:PhysRevLett:05}. One of the two bright states can be selectively excited via a suitable choice of pulse polarization. We neglect the impact of spin-flip in the following, which is typically a good approximation in InAs/GaAs QDs \cite{Narvaez:PhysRevB:06}, where the spin-orbit coupling is weak \cite{Vachon:PhysRevB:09}. Then, due to spin preservation, Coulomb-induced transitions to dark states are suppressed, and we can neglect the influence of dark states on the radiative lifetime of the considered exciton states \cite{Narvaez:PhysRevB:06}. Therefore, we restrict the model of each QD to a two-level system in the following and neglect the spin degree of freedom, as was done, e.g., in Ref. \cite{Nazir:PhysRevB:06}.
The states are denoted by $|0\rangle$ and $|1\rangle$. For the two QDs, this leads to a four-level system with the basis $\{|00\rangle,|01\rangle,|10\rangle,|11\rangle\}$, where $|ij\rangle$ denotes the presence (= 1) or lack (= 0) of an exciton in QD $i$ and $j$, respectively.

The Hamiltonian of the system then reads
\begin{equation}
H=H_e +H_{e-l}+H_{e-e} .
\end{equation}
The noninteracting electrons in effective mass approximation are described by
\begin{equation}
H_e=\sum_{n} \left( \varepsilon_{v,n}a^{\dagger}_{v,n}a_{v,n} + \varepsilon_{c,n}a^{\dagger}_{c,n}a_{c,n} \right),
\end{equation}
with the energy levels $\varepsilon_{\lambda,n}$ of the electrons in the valence ($\lambda=v$) and conduction band ($\lambda=c$) in QD $n$ and the electron creation and annihilation operators $a^{\dagger}_{\lambda,n}$ and $a_{\lambda,n}$, respectively. The Hamiltonian describing the interaction with the light field is given by 
\begin{equation}
H_{e-l}=-\hbar \Omega\sum_{n}(a_{c,n}^{\dagger}a_{v,n}
+a_{v,n}^{\dagger}a_{c,n}
),
\end{equation}
with the Rabi frequency $\Omega =\mathbf{E}(t)\cdot\mathbf{d}_{cv}/\hbar$, 
the classical light field $\mathbf{E}(t)$, and the (real) dipole matrix element 
$\mathbf{d}_{cv}$. The pulse area is defined via the envelope $\Omega_0(t)$ of the pulse $\Omega(t) = \Omega_0(t) \cos(\omega_L t)$, where $\omega_L$ is the pulse frequency:
\begin{equation}
\theta = \int\limits_{-\infty}^\infty dt' \Omega_0(t') .
\end{equation}
The Coulomb interaction between the electrons within the above approximations is described by \cite{Lovett:PhysRevB:03,Danckwerts:PhysRevB:06}

\begin{equation}
H_{e-e} = \sum_{a,b,c,d} V_{\lambda_a \lambda_b \lambda_c \lambda_d}^{1221} a^{\dagger}_{\lambda_{a},1}a^{\dagger}_{\lambda_{b},2}a_{\lambda_{c},2}a_{\lambda_{d},1},
\end{equation}
with the Coulomb matrix elements $V_{abcd}$. When treated in dipole approximation, the Coulomb interaction reduces to two parts represented by a nondiagonal matrix element $V_F=V_{cvvc}^{1221}$ and a diagonal matrix element $V_{11}=V_{cccc}^{1221}$\cite{Danckwerts:PhysRevB:06}.  The nondiagonal matrix element $V_{F}$ couples the two excited single-excitonic states, leading to the well-known effect of F\"orster energy transfer \cite{Forster:AnnPhys:48,Forster::65}. The diagonal element $V_{11}$ shifts the energy of the state $|11\rangle$ and is usually denoted as the biexcitonic shift.

\begin{figure}
  \centering
  \includegraphics*[width=\linewidth]{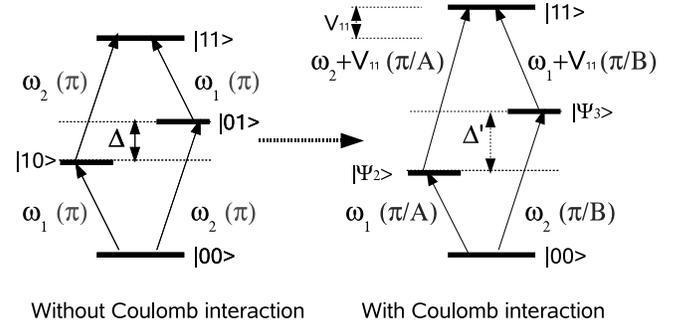}
  \caption{Schematic diagram of the energy levels and transition energies along with the pulse areas required for inversion in the uncoupled and coupled system.
}
  \label{fig:2}
\end{figure}

Defining the matrix elements of the Hamiltonian with respect to the basis $\{|ij\rangle\}$ via $H_{ijkl}= \langle ij|H|kl\rangle$ (cf. Ref. \cite{Danckwerts:PhysRevB:06}), the system Hamiltonian in matrix representation takes the form 
\begin{equation}
  \label{eq:10}
\left(\begin{array}{cccc}
\hbar\omega_{0} &  -\Omega & -\Omega  & 0\\
 -\Omega & \hbar\omega_{0}+\hbar\omega_{1}& V_{F} & -\Omega  \\
 -\Omega & V_{F} &\hbar\omega_{0}+ \hbar\omega_{2} &  -\Omega \\
0 &  -\Omega  & -\Omega &\hbar\omega_{0}+\hbar\omega_{1}+ \hbar\omega_{2}+V_{11}\end{array}\right) ,
\end{equation}
where $\hbar\omega_{0} = \varepsilon_{v,1} + \varepsilon_{v,2}$ denotes the energy of the ground state $|00\rangle$ and $\hbar\omega_{1} = \varepsilon_{c,1} - \varepsilon_{v,1}$ , $\hbar\omega_{2} = \varepsilon_{c,2} - \varepsilon_{v,2}$ are the energies of the single excitons in the two QDs.
 We consider a system with different QD resonance energies $\omega_{2} > \omega_{1}$ with the difference of the single-excitonic energies in the QDs $\Delta =\omega_{2}-\omega_{1} > 0$ [cf. Fig.~\ref{fig:2}(left)]. With a suitable choice of $\Delta$, this allows for selective excitation of the individual QDs  which is necessary for the proposed quantum information schemes. In the following, we refer to the energetically lower QD as the first QD and the energetically higher QD as the second QD.

Starting with this Hamiltonian, we calculate the Heisenberg equations of motion for the density matrix elements of the system in rotating wave approximation \cite{Rossi:RevModPhys:02}. To allow for radiative damping of the system, the interaction of the electrons with the quantized light field, treated in Born-Markov approximation, is included in the calculation. This results in radiative damping terms of the electronic density matrix elements. The full set of equations is discussed in Ref. \cite{Danckwerts:PhysRevB:06}.

The Hamiltonian of the system can be diagonalized with respect to the Coulomb interaction, leading to new eigenstates and eigenvalues:
\begin{align} \label{eq:11}
&|\Psi_{1}\rangle = |00\rangle, &&\lambda_{1}=\hbar\omega_{0} ,\nonumber\\
&|\Psi_{2}\rangle = (c_{1}|10\rangle-c_{2}|01\rangle), &&\lambda_{2}=\hbar\omega_{0}+\hbar\omega_{1}+V_- ,\nonumber\\
&|\Psi_{3}\rangle = (c_{2}|10\rangle+c_{1}|01\rangle), &&\lambda_{3}=\hbar\omega_{0}+\hbar\omega_{1}+V_+ ,\nonumber\\
&|\Psi_{4}\rangle = |11\rangle, &&\lambda_{4}=\hbar\omega_{0}+\hbar\omega_{1}+\hbar\omega_{2}+V_{11} ,
\end{align}
where $V_\pm = \frac{1}{2}(\Delta \pm \sqrt{\Delta^{2}+4V_{F}^{2}})$.
The factors $c_{1}, c_{2}$ depend on $V_F$ via $V_\pm$:
\begin{eqnarray} \label{eq:25}
c_{1} &=& 2 V_+ \left( 4V_{F}^{2}+4V_+^2 \right)^{-\frac{1}{2}},\nonumber\\
c_{2} &=& 2V_{F}(4V_{F}^{2}+4V_+^2)^{-\frac{1}{2}} .
\end{eqnarray}
The new resonance energies of the single excitons are the given by
\begin{eqnarray} 
\label{eq:26}
\hbar\omega_{1}'=\hbar\omega_{1}+V_{-}
,\\
\hbar\omega_{2}'=\hbar\omega_{1}+V_{+}.
\end{eqnarray} 
The energy difference between the two resonances changes from $\Delta$ to 
\begin{eqnarray}
  \label{eq:27}
  \Delta'=
\sqrt{\Delta^{2}+4V_{F}^{2}}.
\end{eqnarray}
In this basis, the Hamiltonian of the system takes the form
\begin{equation}
  \label{eq:10a}
\left(\begin{array}{cccc}
\lambda_{1} &  -\Omega A& -\Omega B & 0\\
 -\Omega A& \lambda_{2}& 0 & -\Omega B \\
 -\Omega B& 0 & \lambda_{3} &  -\Omega A\\
0 &  -\Omega A & -\Omega B& \lambda_{4}\end{array}\right)
\end{equation}
with the renormalization factors $A=c_{1}-c_{2}$ and $B=c_{1}+c_{2}$.

The modification of the energy levels and the transition energies due to the Coulomb interaction is depicted in Fig.~\ref{fig:2}. All transition energies in the four-level system are altered by the F\"orster interaction and the biexcitonic shift. For optical gating, the influence of the Coulomb interaction on the effective Rabi frequencies during optical excitation is of utmost importance: Via the factors $A$ and $B$, the Rabi frequencies now depend on the involved states of the coupled system. 
As an example, we consider the creation of an exciton in the second QD, corresponding to one of the transitions $|00\rangle\rightarrow |\Psi_{3}\rangle$ or $|\Psi_{2}\rangle\rightarrow |11\rangle$. For the first transition, the pulse area $\theta = B^{-1}\pi$ is required, while the second transition requires the pulse area $A^{-1}\pi$. 
This is in contrast to the case of a single, uncoupled dot, where a pulse area of $\pi$ is needed to create full inversion.
This state-dependent renormalization effect plays an important role for the applicability of quantum gates in this system as will be seen in the next section.

Of course, the applicability of the considered system for quantum information processing depends crucially on the size of the physical parameters. For QDs in the low temperature regime, we assume a radiative damping constant of $\gamma$ = (500 ps)$^{-1}$ \cite{Borri:PhysRevLett:01}. The energy difference of the QDs is taken to be  to $10$ meV, which is
well in the range of the inhomogeneous broadening for different quantum dot types \cite{Borri:PhysRevLett:01,Fafard:PhysRevB:94}.
The pulse duration is then strongly constricted, since on the one hand it has to be long enough to allow for spectral selective excitation of the QDs, and on the other hand, it must be short enough to minimize the effects of damping during gating.  It is here taken to be $500$ fs, which corresponds to a spectral pulse width of $2$ meV.
For the implemetation of the CNOT operation, this calls for a biexcitonic shift of a few meV, as will be seen below.

The size of the Coulomb matrix elements $V_{F}$ and $V_{11}$ depends strongly on the material parameters and geometry of the specific QD setup such as the QD material and size, the detuning, and the interdot distance. The values can be as high as $V_{F}/\Delta$ = 0.25 (calculated for self-assembled InAs/GaAs QDs with $\Delta$ = 2  meV \cite{Rozbicki:PhysRevLett:08}) and $V_{11}/\Delta$ = 1.6 (calculated for CdSe quantum dots with $\Delta\approx$ 75 meV under the influence of a static electric field of $100$ kV/cm \cite{Lovett:PhysRevB:03}). On the other hand, very small values can always be obtained by a suitable choice of parameters (for example, 
$V_{F}/\Delta\approx 6\cdot 10^{-3}$ 
in CdSe QDs \cite{Lovett:PhysRevB:03} and $V_{11}/\Delta\approx 1\cdot 10^{-2}$ in GaAs interface QDs \cite{Unold:PhysRevLett:05}).
The coupling constants can be further influenced by external electric fields \cite{Lovett:PhysRevB:03,Biolatti:PhysRevB:02}. 
Instead of focusing on a specific parameter set, we investigate the system for a range of values of the matrix elements known to represent different QD systems.
%

\section{Fidelity of specific quantum operations}

The physical realization of a desired gate will always lead to an
output state with a certain amount of error compared to the ideal result. This
error can be quantified with the fidelity of the process \cite{Nielsen::00}:
the gate fidelity $F$ measures the overlap between the ideal final state and the actual final state which results by applying the gate $G$. If the
result of the ideal gate is a pure state $|\Psi\rangle$ and the actual final state is given by the density matrix $\rho_{G}$, the fidelity is defined as 
\begin{equation}
  \label{eq:12}
  F=(\langle\Psi|\rho_{G}|\Psi\rangle)^{\frac{1}{2}}=({\rm tr}(|\Psi\rangle\langle\Psi|\rho_{G}))^{\frac{1}{2}}.
\end{equation}
If the  result $\rho_{G}$ is identical to the desired pure state, the
fidelity $F=1$. For $F\neq1$, one typically defines the error of the operation as $\delta=1-F^{2}$.
Expanding the pure state $\Psi$ in the basis $\{ |ij\rangle \}$, $|\Psi\rangle =\sum_{ij}c_{ij}|ij\rangle$, Eq.~(\ref{eq:12}) takes the form
  \begin{eqnarray}
    F&=&(\sum_{ijkl}c_{ij}c_{lk}^{*}{\rm tr}(|ij\rangle\langle kl|\rho_{G}))^{\frac{1}{2}}\nonumber \\
&=&(\sum_{ijkl}c_{ij}c_{lk}^{*}\langle a_{i}^{\dagger}a_{j}^{\dagger}a_{k}a_{l}\rangle)^{\frac{1}{2}} .
  \end{eqnarray}
The fidelity is therefore determined by the four-operator expectation values of the system, i.e. the elements of the two-electron density matrix.
For the NOT gate on the first qubit, for example, the fidelity for the input state $|00\rangle$ which is mapped to $|10\rangle$ is just given by $F=\langle a_{c1}^{\dagger}a_{v2}^{\dagger}a_{v2}a_{c1}\rangle^{1/2}$.

Quantum operations on Coulomb-coupled QDs can be realized following the proposition in Ref. \cite{Biolatti:PhysRevB:02}: 
Under resonant coherent excitation of the excitons, the system can be driven from one state to the other, thus realizing gates on the computational states. The Coulomb interaction is used to implement conditional gates, in the following the CNOT gate.
However, the resulting F\"orster interaction mixes the basis states and therefore destroys the original computational basis as seen in Eq.~(\ref{eq:10}).  
As was seen above, the basis in the presence of F\"orster coupling is given by $\{|00\rangle,|\Psi_{2}\rangle,|\Psi_{3}\rangle,|11\rangle\}$ (cf. Ref. \cite{Lovett:PhysRevB:03}). 
Furthermore, the Rabi frequencies for the single-excitonic transitions change from $\Omega$ to $A\Omega$ and $B\Omega$, respectively.
The consequences for potential quantum information processing are the following:
\begin{enumerate}
\item The computational subspace has to be adjusted. Since the states $|10\rangle$  and $|01\rangle$ are no longer optically active when the Coulomb interaction is considered, they cannot be prepared and read out as proposed. Therefore, the states $|\Psi_{2}\rangle$ and $|\Psi_{3}\rangle$ have to be regarded as part of the new computational basis. This 
has to be kept in mind when the system is treated theoretically, especially when calculating the fidelity of quantum operations.
\item F\"orster energy transfer occurs when the system is not prepared in one of its new computational basis states $\{|00\rangle,|\Psi_2\rangle,|\Psi_3\rangle,|11\rangle\}$. Under the influence of decoherence caused by the electron-phonon interaction, it may be irreversible \cite{Rozbicki:PhysRevLett:08}, thus presenting an error that adds to the general error caused by decoherence. Since we neglect electron-phonon coupling here and consider the system prepared in its computational basis, this has no influence in this work.
\item The state-dependent renormalizations of the Rabi frequencies leads to an inevitable error in single-qubit operations since the condition for switching a qubit in one QD depends on the state of the other QD. As discussed below, this error depends on the strength of the F\"orster interaction and adds to the errors caused by decoherence.
\end{enumerate}

\begin{figure}[h]
  \centering
  \includegraphics*[width=\linewidth]{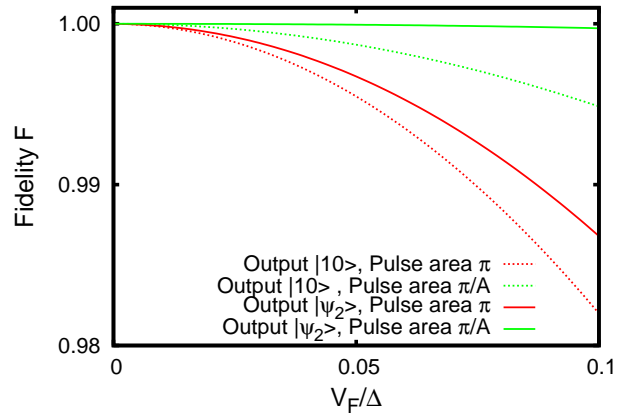}
  \caption{Fidelity of the NOT operation on the first  qubit as a function of the F\"orster coupling constant $V_F$ for the input state $|00\rangle$. The fidelity is shown for an excitation with a $\pi$-pulse (red lines) and  a renormalized ($A^{-1}\pi$-) pulse (green lines), both for the output state $|10\rangle$ (dotted lines) and $|\Psi_{2}\rangle$ (solid lines).}
  \label{fig:3}
\end{figure}

To illustrate the importance of pulse area renormalization and  basis change, we investigate the following situation:  The state $|00\rangle$ is excited by a pulse which is resonant on the first  QD. For an excitation of the uncoupled system with a pulse area $\pi$, this causes a complete inversion of the first QD and therefore corresponds to a NOT operation on the first qubit with the resulting state $|10\rangle$. In the coupled case one would have to adjust the computational subspace and calculate the fidelity with respect to the resulting state $|\Psi_{2}\rangle$ and excite with a pulse area $A^{-1}\pi$, as can be seen in  Fig.~\ref{fig:3}: The fidelity is shown for an excitation with a pulse area $\pi$ (red lines) and the renormalized pulse area $A^{-1} \pi$ (green lines), calculated with the resulting state $|10\rangle$ (solid lines) and $|\Psi_{2}\rangle$ (dotted lines), respectively.
 As expected, the highest fidelity is achieved when working with a renormalized pulse area and within the new computational basis which arises under Coulomb coupling. In this case, the fidelity can be optimized to high values even for large values of the F\"orster coupling $V_F$.

As a consequence, we calculate the performance of the following quantum gates on the computational basis $\{|00\rangle,|\Psi_{2}\rangle,|\Psi_{3}\rangle,|11\rangle\}$ using  pulse areas renormalized with the factors $A^{-1}, B^{-1}$, respectively. The fidelity is usually calculated by averaging over all possible computational input states. In our case, there are important qualitative differences in the performance of the four basis states of the computational subspace. Therefore, we investigate the fidelity for these four basis states separately.



\subsection{NOT gate}

In this section, we investigate the NOT gate on the first qubit of our system.
It is implemented by applying an $A^{-1}\pi$-pulse which is resonant on the first QD (see Table~\ref{table:1} for the corresponding truth table). Since the biexcitonic shift is not important for the realization of the NOT gate, we show the results for $V_{11}=0$. 
\begin{table}[b] \label{table:1}
  \caption{Truth table for qubit gates}
  \begin{tabular}{@{}lllll@{}}
    \hline
     & $|00\rangle$ &  $|\Psi_{2}\rangle$ & $|\Psi_{3}\rangle$ & $|11\rangle$ \\
    \hline
    NOT on Qubit 1 & $|\Psi_{2}\rangle$   & $|00\rangle$   & $|11\rangle$   & $|\Psi_{3}\rangle$      \\
    CNOT  on Qubit 2 & $|00\rangle$   & $|11\rangle$   & $|\Psi_{3}\rangle$   & $|\Psi_{2}\rangle$      \\
    \hline
  \end{tabular}
 \label{truthtable}
\end{table}

\begin{figure}
  \centering
  \includegraphics[width=\linewidth]{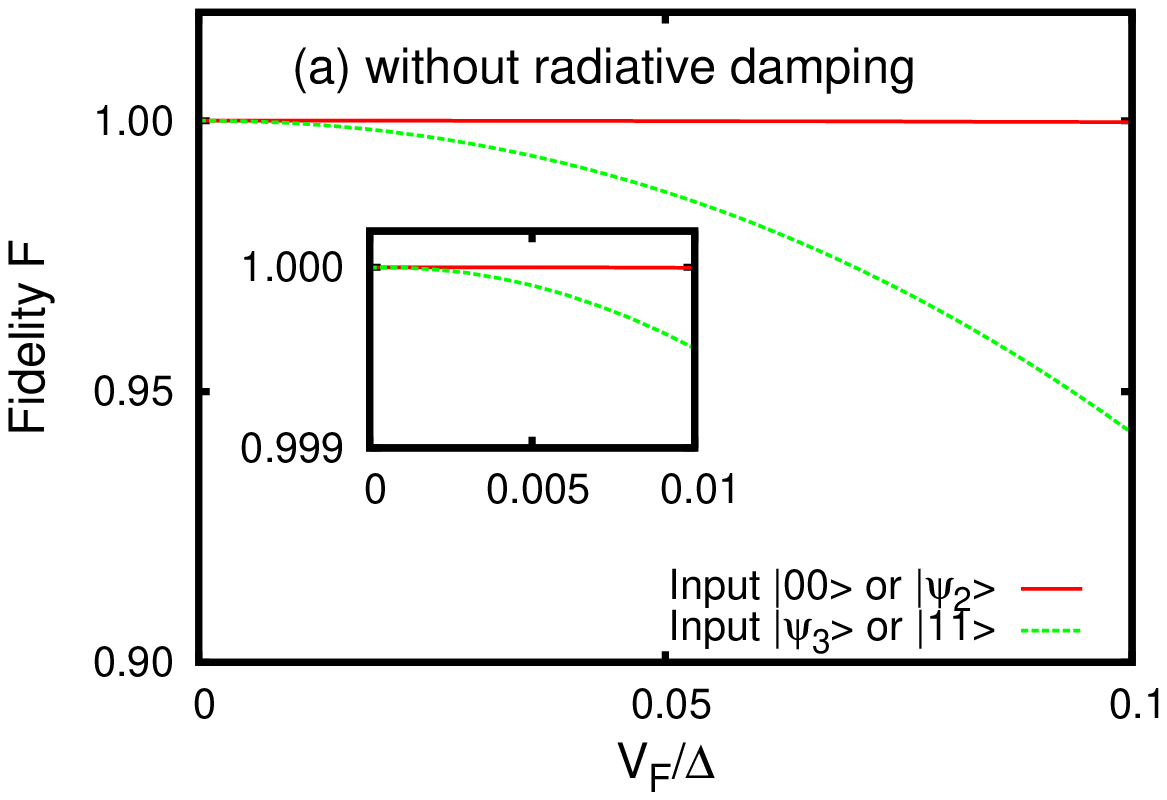}
  \includegraphics[width=\linewidth]{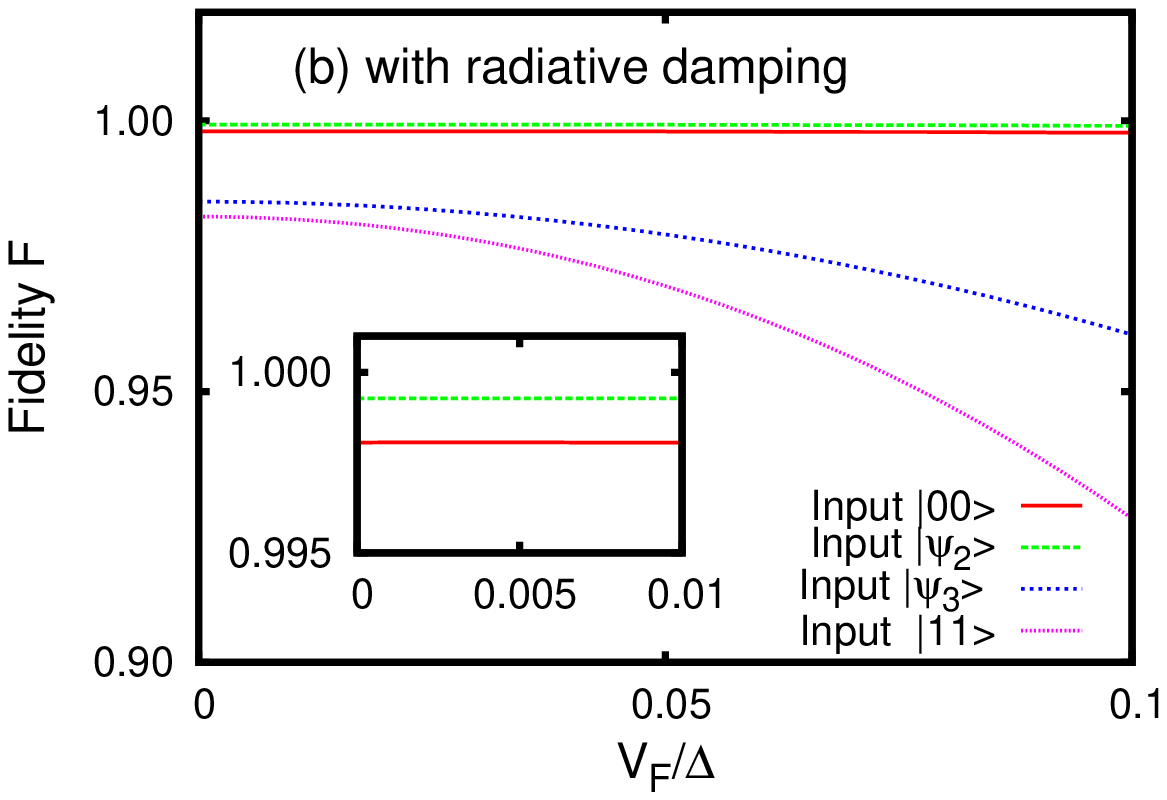}
  \caption{Fidelity of the NOT operation on the first qubit 
as a function of the F\"orster coupling constant $V_F$ for the four input states constituting the computational basis: (a) Without radiative damping, (b) with radiative damping [$\gamma$ = (500 ps)$^{-1}$]. The biexcitonic shift is set to $V_{11}=0$.}
  \label{fig:6}
\end{figure}

In Fig.~\ref{fig:6}(a), the calculated fidelity $F$ is shown for the system without radiative damping. For the states $|00\rangle$ and $|\Psi_2\rangle$, the renormalized pulse area leads to a high fidelity, as is expected from the prior considerations.  Without damping, the states $|00\rangle$ / $|\Psi_2\rangle$ and  $|\Psi_3\rangle$ / $|11\rangle$ show the same fidelity due to the symmetry of excitation and deexcitation.
For the states $|\Psi_3\rangle$ and $|11\rangle$, the renormalization leads to a diminished fidelity. 
This is due to the fact that these states are excited with the ``wrong'' pulse area, as discussed above. 
Thus, the adaptation of the pulse area does not optimize the gate for all possible input states
: On the average, the fidelity is diminished compared to the uncoupled case.  In this way, the F\"orster interaction presents a problem for the proposed quantum information schemes and has to be minimized.

In the inset in Fig. \ref{fig:6}(a), one can see that the error stays in the range of $10^{-4}$ for small values (up to hundreds of $\mu$eV) of $V_F$. 
In Ref. \cite{Lovett:PhysRevB:03}, 
a F\"orster coupling constant of $V_{F}$ = 0.45 meV, corresponding to $V_{F}/\Delta\approx 6\cdot 10^{-3}$ was calculated for a CdSe double QD structure. It was also shown that the F\"orster coupling can be suppressed by applying a static electric field. Since this simultaneously allows an enhancement of the biexcitonic shift \cite{Lovett:PhysRevB:03,Biolatti:PhysRevB:02}, this is one possibility to improve the gate performance of the NOT gate.
It should be noted that there is an additional small fidelity loss with rising coupling strength that cannot be seen in Fig. \ref{fig:6}(a): For $V_{F}/\Delta=0.1$, it adds up to $3\times10^{-4}$. This has a noticeable impact on the generation of Bell states as will be seen in Sec.~\ref{BellStates}.

In a second step, we calculate the performance of the gate under the influence of radiative damping [Fig.~\ref{fig:6}(b)]. 
Now the situation is more complex, since the damping affects the states differently: For $|\Psi_{3}\rangle$ and $|11\rangle$, where both input and output are damped states (see Table~\ref{truthtable}), damping causes a considerable error in the range of $10^{-2}$. The situation is much better for the input states $|\Psi_{2}\rangle$ and $|00\rangle$, where the undamped state $|00\rangle$ is involved either as input or output in the gating process. Here, the error stays in the range of   $10^{-3}$, as can be seen in the inset of Fig.~\ref{fig:6}(b).

\subsection{CNOT gate}

\begin{figure}
   \centering
  \includegraphics[width=\linewidth]{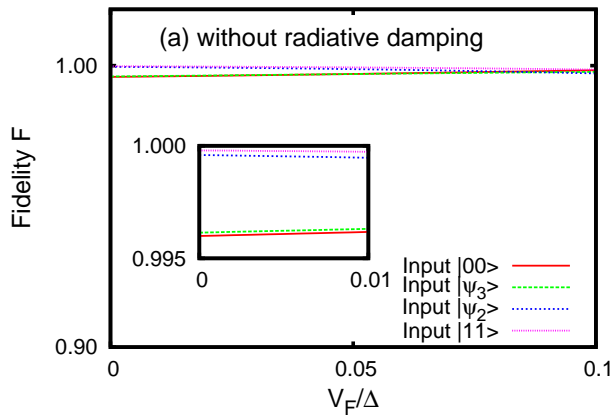}
  \includegraphics[width=\linewidth]{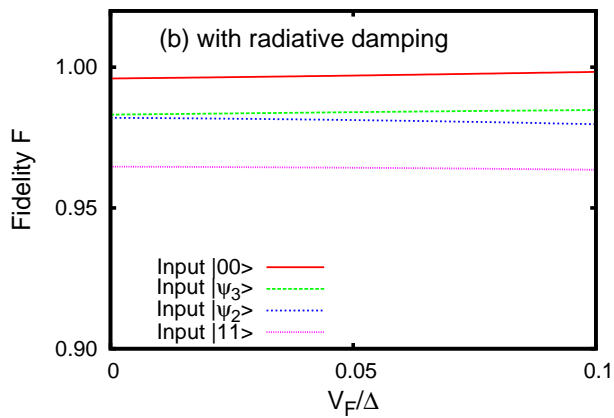}
  \caption{Fidelity of the CNOT operation on the second qubit for the four input basis states as a function of the F\"orster coupling constant $V_F$ (a) without and (b) with radiative damping [$\gamma$ = (500 ps)$^{-1}$] for a biexcitonic shift of $V_{11}$ = 5 meV ($V_{11}/\Delta=0.5$).}
  \label{fig:7}
\end{figure}

The CNOT gate is implemented as suggested in Ref.~\cite{Biolatti:PhysRevLett:00}: The system is excited with an $A^{-1}\pi$-pulse  that is tuned to the exciton-biexciton transition of the second QD. This way, the resonance condition is only fulfilled when the first QD is in its excited state, thus representing a conditional NOT operation on the second qubit (see Table~\ref{table:1} for the truth table of the single-qubit CNOT operation on the second qubit).
In Fig.~\ref{fig:7}, the fidelity of the CNOT operation is shown for a biexcitonic shift of $V_{11}=5$ meV as a function of $V_F$. Without damping [Fig.~\ref{fig:7}(a)], 
the performance is better than that of the NOT gate, since only the two input states $|\Psi_{3}\rangle$ and $|11\rangle$ are affected by the gate. As seen in the truth table, the other two states remain unchanged, and therefore the pulse areas can be optimized for the mapping of $|\Psi_{3}\rangle$ and $|11\rangle$.
In consequence, it does not show the strong dependence on the F\"orster coupling constant $V_{F}$ that is observed for the NOT gate. As can be seen in the inset, the error is of the order of $10^{-3}$ for all input states.
Similar to the NOT gate, the situation changes drastically when damping is considered [see Fig. \ref{fig:7}(b)]: Except for the ground state $|00\rangle$, the fidelity shows a significant reduction in the range of $10^{-2}$, the same order of magnitude as for the NOT gate.
For both cases (with and without damping), there is a small increase of the fidelity for the input states  $|00\rangle$ and $|\Psi_{2}\rangle$ with increasing $V_{F}$. This is due to the corresponding increase of the energy difference of the single-excitonic transitions with $V_{F}$ [see Eq.~(\ref{eq:27})]. 
Even for an energy difference between the two QDs of $10$ meV, there is still a spectral overlap of the $500$ fs Gaussian pulse with the first QD. This overlap decreases when the energy difference increases; therefore, the fidelity increases.
 
\begin{figure}
  \centering
  \includegraphics[width=\linewidth]{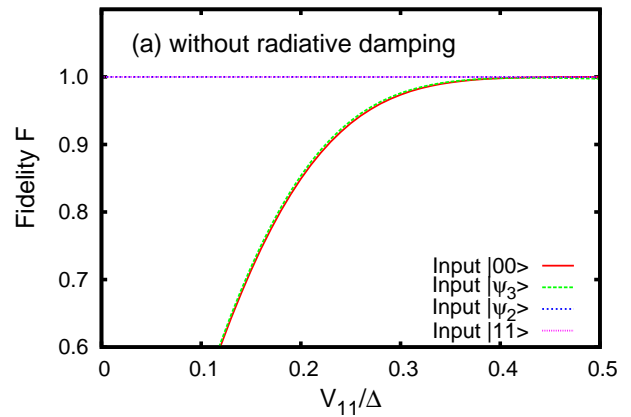}
  \includegraphics[width=\linewidth]{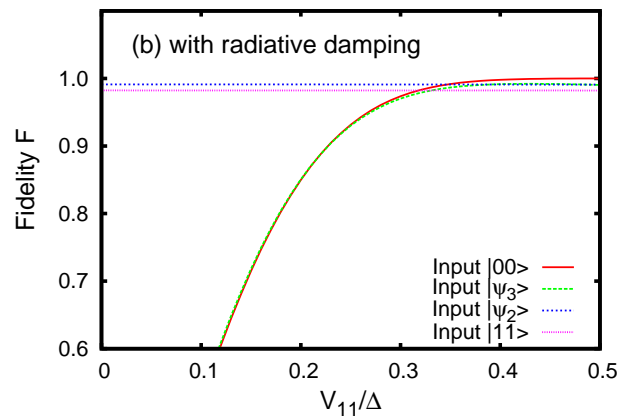}
  \caption{Fidelity of the CNOT operation on the second qubit for the four input basis states as a function of the biexcitonic shift $V_{11}$ (a) without and (b) with radiative damping [$\gamma$ = (500 ps)$^{-1}$]. The fidelity is shown for $V_{F}=0$. 
}
  \label{fig:7a}
\end{figure}

Of course, the performance of the CNOT gate depends crucially on the size of the biexcitonic shift $V_{11}$, 
since the energy shift is the condition for the selective excitation upon which the CNOT gate bases. 
As shown in Fig. \ref{fig:7a} for a vanishing F\"orster coupling $V_F = 0$, for the pulse durations used here (which are restricted by the damping constant of the system), the biexcitonic shift has to be at least $4$ meV to ensure an error of $10^{-3}$ or less. In the presence of damping, this error rate can only be reached for the input state $|00\rangle$; the other input states result in higher errors. The more complex CNOT operation performs better than the NOT gate, especially for high values of $V_{F}$, since, as explained above, it does not show such a strong dependence of the error on $V_{F}$.
As a result, the generation of entangled states works surprisingly well in this system, as will be seen in the next section.

\subsection{\label{BellStates}Bell states}

Bell states are a key ingredient for 
basic quantum information processes such as teleportation \cite{Nielsen::00}. As maximally entangled states, they are also of general interest in quantum information theory. A physical system that is to be used for quantum information processing must provide the possibility to generate and manipulate maximally entangled states.
In our case, the CNOT gate is utilized to create a maximally entangled state in the system, the Bell state $|\Psi_{B}\rangle=\frac{1}{\sqrt{2}}(|00\rangle+|11\rangle)$: One
of the QDs (here, the first QD) is
excited with a $(2A)^{-1} \pi$-pulse, leaving the qubit in an equal
superposition of the ground state 
and $|\Psi_{2}\rangle$. Then, the second QD is excited with an $A^{-1} \pi$-pulse resonant on the biexcitonic resonance, thus selectively exciting
the part of the two-electron state where the first qubit is
already excited.

\begin{figure}
  \centering
  \includegraphics[width=\linewidth]{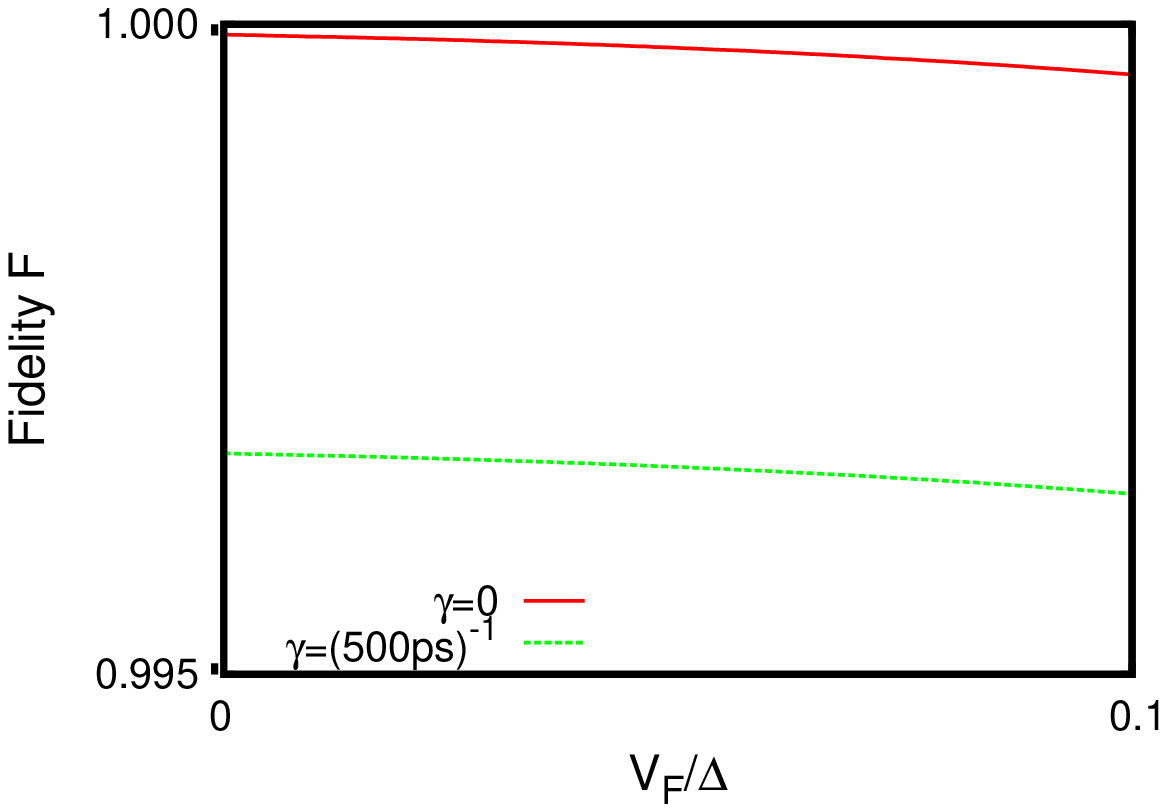}
  \includegraphics[width=\linewidth]{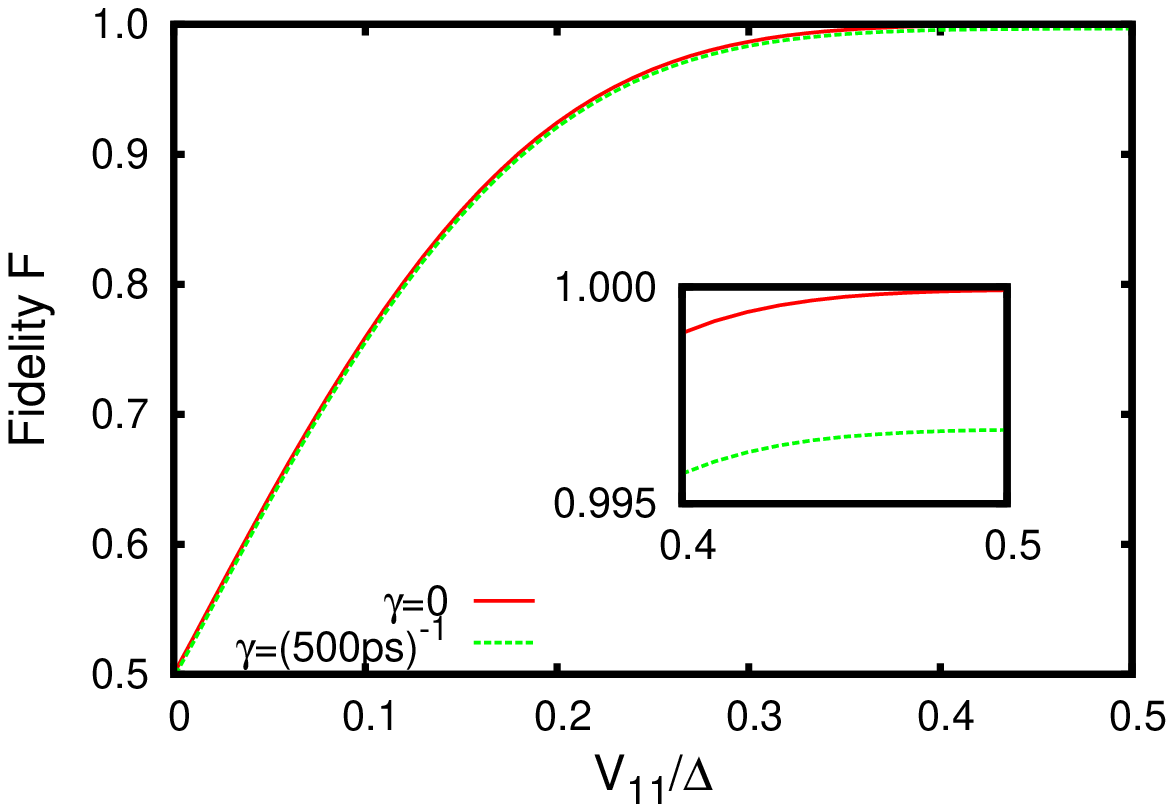}
  \caption{Fidelity for the creation of the Bell state $|\Psi_{B}\rangle=\frac{1}{\sqrt{2}}(|00\rangle+|11\rangle)$ as a function of the Coulomb coupling strength without (red line) and with (green line) radiative damping. Top: the dependence on $V_{F}$ for a biexcitonic shift of $V_{11}= 5$ meV ($V_{11}/\Delta=0.5$). Bottom: the dependence on $V_{11}$ for a F\"orster coupling constant of $V_{F}=0$.}
  \label{fig:8}
\end{figure}

The resulting  fidelity is shown in Fig.~\ref{fig:8}. 
In Fig.~\ref{fig:8}(top), a slight $V_{F}$ dependence of the fidelity can be seen which resembles the one observed for the single-qubit NOT gate.
Figure~\ref{fig:8}(bottom) shows the $V_{11}$ dependence which arises from the underlying CNOT gate. As a combination of the two gates, the generation of Bell states cannot be more efficient than the separate NOT and CNOT gate. However, since a specific input state is used here, the pulse areas are optimized for this input state, and thus a high fidelity can be achieved for this process. Although the errors add up, the value is still acceptable:
Even when radiative damping is taken into account, the error stays in the range of $10^{-3}$. 

\section{Conclusions}

We have investigated the influence of the Coulomb interaction on the applicability of a single- and a two-qubit quantum gate. We find that the F\"orster interaction and radiative damping cause considerable error rates in the fidelity of the gates. While the error caused by the F\"orster interaction can be kept in the acceptable range of $10^{-3}$ by a suitable choice of the system parameters, the fidelity loss due to radiative damping is typically larger and more difficult to deal with. Despite the fidelity loss of the basic gates, the generation of a maximally entangled Bell state is possible with error rates in the range of $10^{-3}$, even in the presence of radiative dephasing.

In general, for using the considered process for successful quantum information processing, one has to think about ways of minimizing the dephasing due to radiative damping, for example by placing the quantum dots in a cavity \cite{Bayer:PhysRevLett:01} or exploiting dark excitons \cite{Narvaez:PhysRevB:06}. Future work should then also focus on the error caused by the interaction with phonons.





\begin{acknowledgements}
We thank A. Grodecka-Grad for fruitful discussion.
\end{acknowledgements}

%
%


\providecommand{\WileyBibTextsc}{}
\let\textsc\WileyBibTextsc
\providecommand{\othercit}{}
\providecommand{\jr}[1]{#1}
\providecommand{\etal}{~et~al.}

\end{document}